\documentclass[11pt]{article}
\usepackage{amsfonts}
\usepackage{amssymb}
\usepackage{amsmath}
\usepackage{latexsym}
\usepackage{graphicx}
\usepackage[english]{babel}

\input epsf

\newcommand\be{\begin{equation}}
\newcommand\bea{\begin{eqnarray}}
\newcommand\ee{\end{equation}}
\newcommand\eea{\end{eqnarray}}
\newcommand\h{\frac{1}{2}}
\newcommand\Regge{\alpha'}
\newcommand{\bdm}{\begin{displaymath}}
\newcommand{\edm}{\end{displaymath}}

\newcommand{\f}[2]{\frac{#1}{#2}}
\newcommand{\bref}[1]{(\ref{#1})}
\newcommand\p{\partial}
\newcommand \nn{\nonumber\\ }

\renewcommand\r{\rightarrow}

\begin{document}

\begin{flushright}
\end{flushright}
\vspace{20mm}
\begin{center}
{\LARGE   Non-extremal fuzzballs and ergoregion emission}\\
\vspace{18mm}
{\bf Borun D. Chowdhury}\footnote{E-mail: {\tt borundev@pacific.mps.ohio-state.edu}.}, 
{\bf and Samir D. Mathur}\footnote{E-mail: {\tt mathur@mps.ohio-state.edu}.}\\
\vspace{8mm}
Department of Physics,\\ The Ohio State University,\\ Columbus,
Ohio, USA 43210\\
\vspace{4mm}
\end{center}
\vspace{10mm}
\thispagestyle{empty}
\begin{abstract}

In the traditional picture of black holes Hawking radiation is created  by pair creation from the vacuum at the horizon. In the fuzzball proposal, individual microstates do not have a horizon with the `vacuum' state  in its vicinity. For a special family of non-extremal microstates it was recently found that emission occurs due to pair creation in an ergoregion, rather than at a horizon. In this paper we extend this result to a slightly larger class of microstates, again finding exact agreement between the emission in the gravity picture and the CFT dual. We  write down an expression for emission from geometries with ergoregions,  in terms of the leading falloff behavior of the wavefunctions in the fuzzball region. Finally, we describe another family of nonextremal microstates and find their ergoregion.

\end{abstract}
\newpage
\setcounter{page}{1}

\section{Introduction}
\setcounter{equation}{0}

How does information come out of a black hole? In recent years we have learnt that the internal structure of black holes is different from the traditionally assumed structure. For extremal holes the traditional picture has an infinitely long throat ending in a horizon, and a singularity inside the horizon. But when we construct the states of the hole in the full string theory we find `fuzzballs'; the throat is long but not infinite, and ends in a `quantum fuzz' rather than a horizon \cite{lm4,lm5,fuzzball2}. Which of the $e^{S_{bek}}$ states of the hole we have determines the details of the `fuzz', and quanta falling down the throat can be absorbed, mixed with the fuzz, and ultimately re-emitted, without loss of unitarity.
The 2-charge extremal hole has been quite thoroughly understood\cite{lm4,lm5,lmm,st}, and there are a large class of states known for the 3-charge and 4-charge extremal holes \cite{mss,gms1,gms2,lunin,bena,gimon,giustoRef}. All states constructed so far agree with the fuzzball picture, and not with the traditional picture of the hole.  

Extremal holes do not Hawking radiate, and if we want to see how information comes out in Hawking radiation then we have to look at nonextremal holes. A family of nonextremal microstates was constructed in \cite{ross}. In \cite{myers}  a subfamily of these states were examined and it was found that they radiated energy by `ergoregion instability'. Finally, in \cite{cm1,cm2} it was shown that this instability radiation was exactly the `Hawking radiation' expected from these special microstates. This was done by looking at the emission from the states of the dual CFT.  The process that reproduces the gross properties of Hawking radiation from `thermal' CFT states \cite{radiation} turns out to give exactly the `instability radiation' when applied to the special CFT microstates.

In this paper we wish to extend these results about nonextremal microstates in some small ways:

(a) The computations of instability in \cite{myers} and its matching to Hawking radiation rates in \cite{cm1,cm2} were carried out   for `maximally rotating states'. The geometries of \cite{ross} also cover microstates with lower rotation; these geometries have conical defects in their interior and correspond in the CFT to having `multiwound component strings'. We extend the computations of \cite{myers} and \cite{cm1,cm2} to cover these
more general microstates. Thus we solve the wave equation for a scalar in these geometries, and find the instability frequencies. We then look at the emission process in the dual CFT state, and show that both the real and imaginary parts of the frequencies are exactly reproduced. 

(b) The above mentioned microstates had a $U(1)\times U(1)$ symmetry. We can consider more general microstates which are stationary geometries with an ergoregion. We cannot explicitly solve the scalar wave equation in these general geometries. But the kind of geometry that is of interest to us can  be separated into a `cap region' (which has all the complications of the microstate) and an $AdS$ region where the solution of the wave equation has a universal form. Solving the wave-equation in the `cap'  leads to the coefficients of the two allowed solutions in the $AdS$ region. We find an expression for the real and imaginary parts of the ergoregion instability frequencies in terms of the ratio $F(\omega)$ of these two coefficients.  

(c) We find another family of non-extremal microstates for the D1-D5 system which are stationary geometries. We then find the ergoregion for these geometries. We observe that in the limit of large non-extremality, the ergoregions of these geometries and those of \cite{ross} tend to the same limit. 

\bigskip

We conclude with a short discussion of the significance of microstates with ergoregions for the problem of black hole radiation.

\section{The microstate geometry}\label{section2}
\setcounter{equation}{0}

Let us start by recalling the microstate geometries that we will consider. These geometries were constructed in \cite{ross}. We describe the geometries, and then explain the limits of parameters that we will take in carrying out our computations. 

\subsection{The supergravity solution}

To construct the geometries in \cite{ross} one starts with supergravity solutions for arbitrary charges and angular momenta \cite{cveticyoum,cveticlarsen} and then chooses the parameters such that in the dual CFT there is a unique state with those properties. This procedure gives a geometry dual to the CFT state. This geometry turns out to not have a horizon, and is also regular except for possible conical defects. This procedure was used to construct 2-charge extremal geometries in \cite{balmm},  3-charge extremal geometries in \cite{gms1,gms2} and  a family of non-extremal geometries in \cite{ross}. In \cite{cm1,cm2} we were concerned with geometries from \cite{ross} which did not have a conical defect while in this paper we will use geometries which do have conical defects.

Take type IIB string theory, and compactify 10-dimensional spacetime as
\be
M_{9,1}\rightarrow M_{4,1}\times T^4\times S^1
\ee
The volume of $T^4$ is $(2\pi)^4 V$ and the length of $S^1$ is $(2\pi) R$. The $T^4$ is described by coordinates $z_i$ and the $S^1$ by a coordinate $y$. The noncompact $M_{4,1}$ is described by a time coordinate $t$, a radial coordinate $r$, and  angular $S^3$  coordinates $\theta, \psi, \phi$. The solution will have angular momenta along $\psi, \phi$, called $J_\psi, J_\phi$, captured by two parameters $a_1, a_2$.
The solutions will carry three kinds of charges. We have $n_1$ units of D1 charge along $S^1$, $n_5$ units of D5 charge wrapped on $T^4\times S^1$, and $n_p$ units of momentum charge P along $S^1$. 
These charges  will be described in the solution by three parameters $\delta_1, \delta_5, \delta_p$. 

In this paper we will look at  states where the P charge is zero ($\delta_i=n_p=0$), since this case will suffice to bring out the observations that we wish to make. It turns out that $n_p=0$ implies that one of the angular momenta vanish: $J_\phi=0$. The resulting geometries are (in the string frame)
\begin{eqnarray} \label{2charge}
ds^2&=&-\frac{f-M}{\sqrt{\tilde{H}_{1} \tilde{H}_{5}}}
dt^2+\frac{f}{\sqrt{\tilde{H}_{1} \tilde{H}_{5}}}dy^2+\sqrt{\tilde{H}_{1} \tilde{H}_{5}}
\left(\frac{ dr^2}{ r^2+a_{1}^2 - M}
+d\theta^2 \right)\nonumber \\ 
&&+\left( \sqrt{\tilde{H}_{1}
\tilde{H}_{5}} + a_1^2 \frac{( \tilde{H}_{1} + \tilde{H}_{5}
-f+M) \cos^2\theta}{\sqrt{\tilde{H}_{1} \tilde{H}_{5}}}  \right) \cos^2
\theta d \psi^2 \nonumber \\ 
&& +\left( \sqrt{\tilde{H}_{1}
\tilde{H}_{5}} -a_1^2 \frac{(\tilde{H}_{1} + \tilde{H}_{5}
-f) \sin^2\theta}{\sqrt{\tilde{H}_{1} \tilde{H}_{5}}}\right) \sin^2
\theta d \phi^2 \nonumber \\ 
&&+ \frac{2M \cos^2 \theta}{\sqrt{\tilde{H}_{1} \tilde{H}_{5}}}(a_1
c_1 c_5 ) dt  d\psi +\frac{2M \sin^2 \theta}{\sqrt{\tilde{H}_{1} \tilde{H}_{5}}}(a_1
s_1 s_5 ) dyd\phi + \sqrt{\frac{\tilde{H}_1}{\tilde{H}_5}}\sum_{i=1}^4
dz_i^2 \label{Eqn:metric}
\end{eqnarray}
where
\be
c_i = \cosh \delta_i, \quad s_i=\sinh \delta_i
\ee
\begin{eqnarray} 
\tilde{H}_{i}=f+M\sinh^2\delta_i, \quad
f=r^2+a_1^2\sin^2\theta, \label{Def:FandH}
\end{eqnarray}
The D1 and D5 charges of the solution produce a RR 2-form gauge field. The RR 2-form gauge field and the dilaton are given in  \cite{gms1,gms2,ross}. The mass of the system is given by
\be
M_{ADM} = \f{\pi M}{4 G^{(5)}} (s_1^2+ s_5^2 + \f{3}{2}) \label{Eqn:Madm}
\ee
The angular momenta are given by
\bea
J_\psi &=& -  \f{\pi M}{4 G^{(5)}} a_1 c_1 c_5   \\
J_\phi &=& 0  \label{angMom}
\eea
It is convenient to define
\be 
Q_1=M\sinh\delta_1\cosh\delta_1, ~~Q_5=M\sinh\delta_5\cosh\delta_5
\label{qdef}
\ee
The integer charges of the solution are related to the $Q_i$ through
\bea
Q_1&=& \frac{g \Regge^3}{V} n_1 \nonumber \\ 
Q_5 &=& g \Regge n_5  \label{q1q5}
\eea
We must further choose
\be
\gamma \f{M s_1 s_5}{\sqrt{a_1^2-M}}=R, \qquad  \f{a_1}{\sqrt{a_1^2 - M}} =m\label{Eqn:Smooth}
\ee
Here $R$ is the radius of the $S^1$. The parameter $m$ is an integer; larger values of $m$ give more non-extremality to the solution while keeping the charges fixed. The parameter $\gamma$ has the form
\be
\gamma={1\over k}, ~~\in \mathbb Z \label{Eqn:GammaAndK}
\ee
The choice (\ref{Eqn:Smooth}) gives geometries that are regular upto possible conical defects.
In the computations of \cite{myers,cm1} the value $\gamma=1$ had been taken. This value makes the geometries completely smooth everywhere (no conical defects). One of the goals of this paper is to extend the earlier computations to the case $\gamma\ne 1$, where we {\it do} have an order $k$ conical defect.  

In our work below we will ignore the torus $T^4$ since none of our variables depend on the torus coordinates. We will work with the 6-d Einstein metric unless otherwise mentioned. It turns out that this metric is the same as (\ref{Eqn:metric}) with the torus contribution discarded.

\subsection{The large R limit}

As explained in \cite{cm1,cm2}, if we want to relate our computations to a dual CFT description then we need to have a large AdS type region in our geometry. Such a region is obtained if we let the radius $R$ of the $S^1$ be large. 
The large R limit is defined by
\be
\epsilon \equiv \f{\sqrt{Q_1Q_5}}{R^2} \ll  1 \label{Def:epsilon}
\ee
We would now like to express the parameters of the solution (\ref{Eqn:metric}) in a way that manifests their behavior in this large $R$ limit. 
From \bref{q1q5}  we can see that $Q_1,Q_5$ do not depend on $R$. In this large $R$ limit we will have $M\ll Q$ and 
\be
s_1 \approx c_1, \quad s_5\approx c_5
\label{firsteq}
\ee
which gives with \bref{qdef}
\be
M s_1^2 =Q_1, \qquad Ms_5^2 = Q_5 \label{LargeRQ}
\ee
We will assume that $Q_1$ and $Q_5$ are of the same order. We see that 
\be
M=(m^2-1) \f{Q_1 Q_5}{(kR)^2}  \label{LargeRM}
\ee
and we get
\be
a_1 = m \f{\sqrt{Q_1 Q_5}}{(k R)}  \label{LargeRa}
\ee
The mass of the extremal D1-D5 system is
\be
M_{extremal} = \f{\pi M}{4 G^{(5)}} (s_1^2+ s_5^2+1)
\ee
This gives the mass above extremality
\be
\Delta M_{ADM} =  \f{\pi M}{8 G^{(5)}}
\ee
Using
\be
16 \pi G^{(10)} = (2\pi)^7 g^2 \alpha^4= 16 \pi G^{(5)} (2 \pi R) (2\pi)^4 V
\ee
and \bref{q1q5} we get
\be
16 \pi G^{(5)}  = (2\pi)^2 \f{Q_1 Q_5}{R n_1 n_5}
\ee
This gives 
\be
\Delta M_{ADM} = \f{n_1 n_5}{2k^2 R} (m^2-1)  \label{Eqn:DeltaM}
\ee
Using  \bref{angMom}, \bref{firsteq}, \bref{LargeRQ} and \bref{LargeRa} we see that the angular momenta are
\bea
J_\psi =- \f{m}{k} n_1 n_5, \qquad J_\phi=0 \label{Eqn:JInt}
\eea
Furthermore with these expressions for $M$ and $a_1$  we get from \bref{Def:FandH}
\bea
f&=&r^2 + m^2  \gamma^2 \f{Q_1Q_5}{R^2} \sin^2 \theta \nonumber \\
\tilde{H}_i &=& r^2+ Q_i \label{Eqn:FandHLargeR}
\eea

\subsection{The inner and outer regions}

In the large $R$ limit we can separate the geometry into two regions: an `inner region' which is an $AdS$ type geometry and an outer region which is essentially flat space. These two regions are connected by a region around $r\sim (Q_1Q_5)^{\f{1}{4}}$ which we will call the `neck'. 

\subsubsection{Inner Region: $r^2  \ll \sqrt{Q_1 Q_5}$}

In this region 
\bea
f&=&r^2 + m^2 \gamma^2  \f{Q_1Q_5}{R^2} \sin^2 \theta \nonumber \\
\tilde{H}_i &=& Q_i 
\eea
The metric takes a simple form in terms of the coordinates
\bea
\tau \equiv \f{t}{R}, \qquad \varphi \equiv \f{y}{R}, \qquad \rho \equiv \f{r R}{\sqrt{Q_1 Q_5}} \label{Def:AdSCood}
\eea
In these coordinates the  metric in the inner region is
\bea
ds^2 &=& \sqrt{Q_1 Q_5} \Bigg[ \Big (-( \rho^2  + \gamma^2) d\tau^2 + \f{d \rho^2}{( \rho^2+ \gamma^2)} + \rho^2 d\varphi^2 \Big ) \nonumber \\
&& + \Big ( d \theta^2 + \cos^2 \theta ( d \psi + m \gamma d \tau)^2 + \sin^2 \theta (d \phi + m \gamma d\varphi)^2 \Big) \Bigg]
\label{metricinner}
\eea
This geometry {\it locally} has the form of $AdS_3$ with an $S^3$ fibered over the $AdS_3$. As mentioned before, there is a conical defect along the curve $\rho=0, \theta={\pi\over 2}$. At this curve we get an ALE singularity of the type which arises when the centers of $k={1\over \gamma}$ KK monopoles are brought together. The fibration is characterized by $\gamma$ and the integer $m$. The $AdS_3$ and the $S^3$ each have curvature radius $(Q_1 Q_5)^\f{1}{4}$. 

The condition defining the inner region  $0<r\ll  (Q_1 Q_5)^\f{1}{4}$ is equivalent to $0< \rho\ll  \f{R}{(Q_1 Q_5)^\f{1}{4}}$. In the large $R$ limit the radial coordinate of the $AdS$ region extends over `many many curvature radii' before we reach the `neck'. Thus we have a large $AdS$ region and a good description in terms of a dual CFT.

\subsubsection{The Outer Region: $r^2 \gg \sqrt{Q_1Q_5}$}

For our purposes it will be adequate to approximate the metric in this region by its leading approximation which is flat spacetime:
\be
ds^2 = - dt^2 + dy^2 + dr^2 + r^2 d\Omega_3^2 
\label{flat}
\ee

\subsection{The ergoregion}

Finally, let us compute the ergoregion for this geometry. This ergoregion is of interest to us because the  emission of energy from  the microstate will occur by pair creation due to the presence of this ergoregion. One member of the pair will stay in the ergoregion, and the other member will be radiated to infinity.

The ergoregion for the geometry in \bref{metricinner} is given by the region where $g_{\tau \tau} >0$ and is thus the region given by
\be
-(\rho^2+ \gamma^2) + m^2 \gamma^2 \cos^2 \theta >0
\ee
i.e.
\be
0 < \rho < \gamma \sqrt{ m^2 \cos^2 \theta-1} \label{Eqn:RossErgoregion}
\ee

\section{The classical instability}

The geometries considered above are non-extremal and have an instability leading to energy being radiated to infinity. In this section we compute the emission of a scalar field from the above discussed geometries, by solving the scalar wave equation in those geometries. For the case $\gamma=1$ this computation was done in \cite{myers}; here we redo the computation in full since we need the result for $\gamma\ne 1$. We will follow the steps of \cite{cm1,cm2}, where a slightly simplified version of the computation of \cite{myers} was presented.

To see the classical instability  we need to solve the scalar wave equation  with purely outgoing boundary conditions. As was shown in \cite{myers,cm1,cm2}, purely outgoing boundary conditions give us two solutions, one of which increases exponentially in time and the other decreases exponentially in time. We will be interested in the former. The details of the derivation can be found in \cite{ross,myers,cm1,cm2};  here we sketch the main steps.

The scalar wave equation is
\be
\square \Phi=0
\ee
It will be convenient to use the radial coordinate
\be
x=\rho^2=\f{r^2 R^2}{Q_1 Q_5}
\ee
where we have used \bref{Def:AdSCood}.
The variables can be separated  \cite{ross,myers,cveticlarsen} and so we take the ansatz
\be
\Phi = e^{- i \omega t + i  m_\psi \psi} \chi(\theta) h(r)
\ee
The radial part of the wave equation is found to be
\be
4 \partial_x(x(x+ \gamma^2) \partial_x h) + ( \kappa^2 x + (1- \nu^2) + \f{\xi^2}{x +\gamma^2} )h=0
\ee
where
\bea
\xi &=& \omega R + \gamma~ m m_\psi \nn
\kappa &=& \omega^2 \f{Q_1 Q_5}{R^2} \label{Def:XiAndKappa} \nn
\nu &\approx& l+1+ O(\epsilon^2)
\eea
The correction $O(\epsilon^2)$ to $\nu$ is given in detail in \cite{cm1} ($\epsilon$ is defined in \bref{Def:epsilon}). The factor $1-\nu^2$ arises from the angular part of the Laplacian.  In the large $R$ limit which we take at the end the $O(\epsilon^2)$ correction  goes to zero, but for now we keep it as a regulator which will shift the arguments of $\Gamma$ functions away from poles.  
 
In the inner region this equation is approximated by 
\be
4 \partial_x(x(x+ \gamma^2) \partial_x h) + (  (1- \nu^2) + \f{ \xi^2}{x +\gamma^2} )h=0
\ee
while in the outer region we get
\be
4 \partial_x(x^2  \partial_x h) + ( \kappa^2 x + (1- \nu^2)  )h=0
\ee
For the inner region the solution which is regular at the origin has the form
\be
h_{in} = (x + \gamma^2)^{\f{\xi }{2 \gamma}} \phantom{a}_2 F_1(\h(1+\f{\xi}{\gamma} - \nu) , \h(1+ \f{\xi}{\gamma} + \nu) ,1 , -\f{x}{\gamma^2} )
\ee
For large $x$ this solution behaves as 
\bea
h_{in+} &=& \gamma^{ \f{\xi}{\gamma}} \Big[ \f{\Gamma(\nu)}{\Gamma(\h(1+ \f{\xi}{\gamma} + \nu))\Gamma(\h(1- \f{\xi}{\gamma} + \nu))} (\f{x}{\gamma^2})^{-\h(1-\nu)} \nn
&& ~~~~~~~~~~~~~~+\f{\Gamma(-\nu)}{\Gamma(\h(1+ \f{\xi}{\gamma} -\nu))\Gamma(\h(1- \f{\xi}{\gamma} -\nu))} (\f{x}{\gamma^2})^{-\h(1+\nu)}  \Big]
\eea
The outer region gives the solution
\be
h_{out} = \f{1}{\sqrt{x}} \left[ C_1 J_{\nu} (\kappa \sqrt{x}) + C_2 J_{-\nu} ( \kappa \sqrt{x}) \right]
\ee
Matching the inner region solution to the outer region solution (with purely outgoing boundary conditions) gives
\be
-e^{-i \pi  \nu} \f{\Gamma(1-\nu)}{\Gamma(1+ \nu)} \left( \f{\gamma \kappa}{2 } \right)^{2 \nu}  = \f{\Gamma(\nu)}{\Gamma(-\nu)} \f{\Gamma(\h(1- \nu + \f{\xi}{\gamma})) \Gamma(\h(1- \nu - \f{\xi}{\gamma})) }{\Gamma(\h(1+ \nu + \f{\xi}{\gamma})) \Gamma(\h(1+\nu - \f{\xi}{\gamma})) }
\ee
This transcendental equation was solved in \cite{myers}, and their solution was rederived  in a slightly different manner in  \cite{cm1,cm2}. The details of the solution method can be found in those references. We note that out of the different possible solutions  we have to take the ones with $\omega_I>0$ as these are the ones which give the unstable modes we are seeking. The solutions to the transcendental equation involve an arbitrary integer $\hat N$. For simplicity we consider only the lowest allowed value $\hat N=0$; this corresponds to the highest allowed emission energy for a given angular harmonic. Then the emission is described by
\be
\omega_R = \f{\gamma}{ R}( -l -2 - m_\psi  m), \qquad \omega_I = \f{\gamma }{R} \f{2 \pi}{(l!)^2} \left ( \f{\gamma \kappa }{2} \right)^{2(l+1)} \label{Eqn:GravOmega}
\ee
Here $\omega_R$ gives the energy of the radiated scalar quanta, and $\omega_I$  gives the rate of growth of the instability. Our goal in the present paper will be to understand the dependence on the parameter $\gamma=\f{1}{k}$. We see that $\gamma$ appears as a simple prefactor in $\omega_R$, and in $\omega_I$ the power of $\gamma$ depends on the angular harmonic $l$.

\section{The CFT description for the state and for the emitted scalar} 
\subsection{The D1-D5 CFT}

In this section we compute the emission from the CFT states dual to the geometries described in section \ref{section2}. The computation will parallel the computation in \cite{cm1} for the case $k=1$. Thus in many places we will refer to \cite{cm1} for  details; the main issue here is to see where the factors of $k$ appear, and to note at the end that they appear exactly as they do in the gravity computation of emission  (\ref{Eqn:GravOmega}).

\subsection{The CFT state}

We first describe the CFT, then the way we construct the microstates of interest. 

\subsubsection{The orbifold CFT}

We take IIB string theory compactified to $M_{4,1}\times S^1\times
T^4$.  Let $y$ be the coordinate along $S^1$ with
\be
0\le y<2\pi R
\ee
The $T^4$ is described by 4 coordinates $z_1, z_2, z_3, z_4$, and
the noncompact space is spanned by $t, x_1, x_2, x_3, x_4$.  We
wrap
$n_1$ D1 branes on $S^1$, and
$n_5$ D5 branes on
$S^1\times T^4$. Let 
\be
N=n_1n_5
\ee
The bound state of these branes
is described by a 1+1 dimensional sigma model, with base space
$(y,t)$ and target space a deformation of the orbifold
$(T^4)^N/S_N$ (the symmetric product of $N$ copies of $T^4$).  The
CFT has ${\cal N}=4$ supersymmetry, and a moduli space which
preserves this supersymmetry. It is conjectured that in this
moduli space we have an `orbifold point' where the target space is
just the orbifold
$(T^4)^N/S_N$ \cite{sw}.

The rotational symmetry of the noncompact directions $x_1\dots
x_4$ gives a symmetry
$so(4)\approx su(2)_L\times su(2)_R$, which is the R symmetry group of
the CFT.

The CFT with target space just one copy of $T^4$ is described by 4
real bosons
$X^1$,  $X^2$,  $X^3$,  $X^4$ (which arise from the 4 directions $z_1,
z_2,
z_3, z_4$), 4 real left moving fermions $\psi^1, \psi^2, \psi^3,
\psi^4$ and 4 right moving fermions $\bar\psi^1, \bar\psi^2,
\bar\psi^3, \bar\psi^4$. The central charge is $c=6$.  Note that the fermions can be either periodic (Ramond sector R) or antiperiodic (Neveu-Schwarz sector NS) as we go around a circle. We can take linear combinations of the fermions $\psi^i$ to get fermions with definite azimuthal quantum number $j$ under $su(2)_L$ \cite{cm1}. We get left moving fermions $\psi,\tilde\psi$ with $j={1\over 2}$, and their conjugates have $j=-{1\over 2}$. The right fermions
$\bar\psi^i$ similarly give $\bar\psi, \tilde{\bar\psi}$ with $\bar j={1\over 2}$   The overall `charge' of a state will be specified by 
two quantum numbers $(j, \bar j)=(j^3_L, j^3_R)$. 

The complete
theory with target space $(T^4)^N/S_N$ has $N$ copies of this
$c=6$ CFT, with states that are symmetrized between the $N$
copies. The orbifolding also generates `twist' sectors, which are
created by twist operators $\sigma_k$. A detailed construction
of the twist operators is given in  \cite{lm1, lm2}, but we summarize
here the properties that will be relevant to us.

Suppose the CFT lives on a spatial circle of length $L$. The twist operator of order $k$ links together $k$ copies of the $c=6$
CFT so that the $X^i, \psi^i$ act as free fields living on a circle
of length $kL$. 
Recall that an operator is called a chiral primary if its  charge $j$ equals its dimension $h$, and an anti-chiral primary if $j=-h$. By adding a suitable charge to the twist operator $\sigma_k$ we can make chiral and anti-chiral primaries. The operator of interest to us is the anti-chiral primary operator
\be
\tilde \sigma_k^{--}: ~~h=\bar h={k-1\over 2}, ~~~j= \bar j=-{k-1\over 2}
\label{operator}
\ee
For a detailed construction of such operators, see\cite{lm2}. 

\begin{figure}[htbp] 
 \begin{center}\hspace{-1truecm}
  \includegraphics[width=2in]{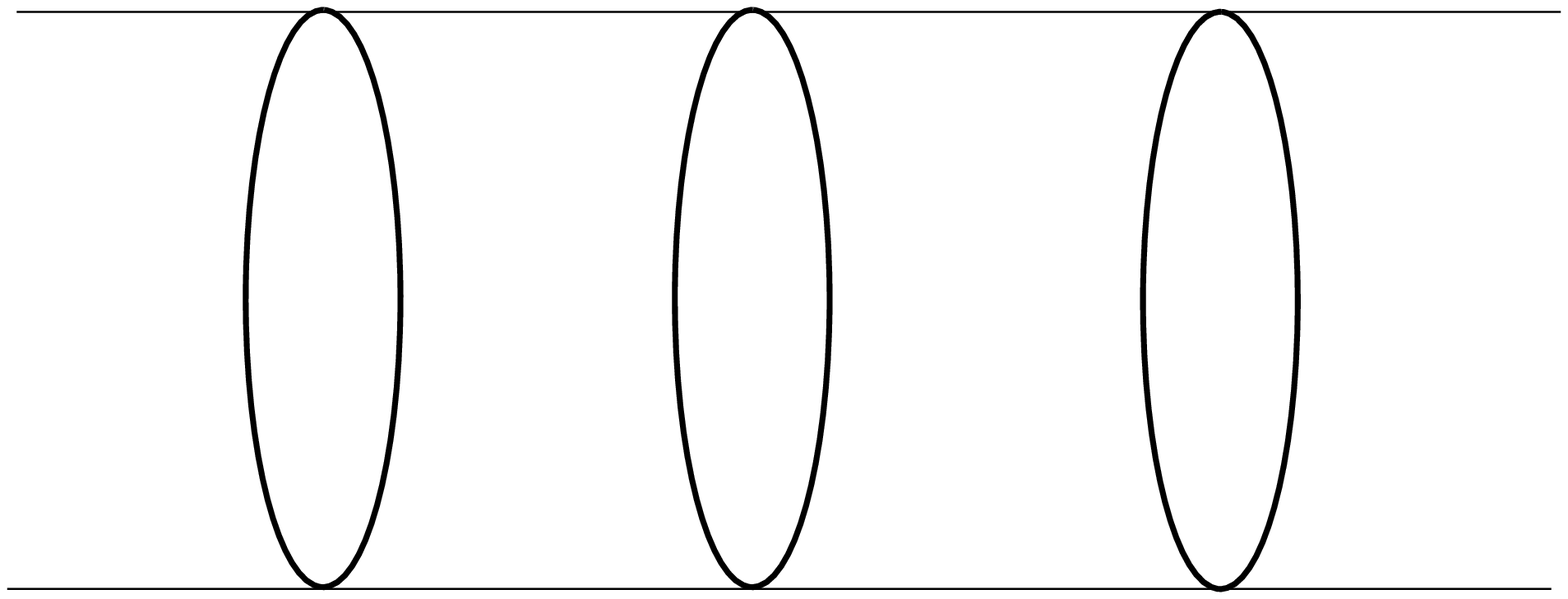} \hspace{1truecm}
    \includegraphics[width=2in]{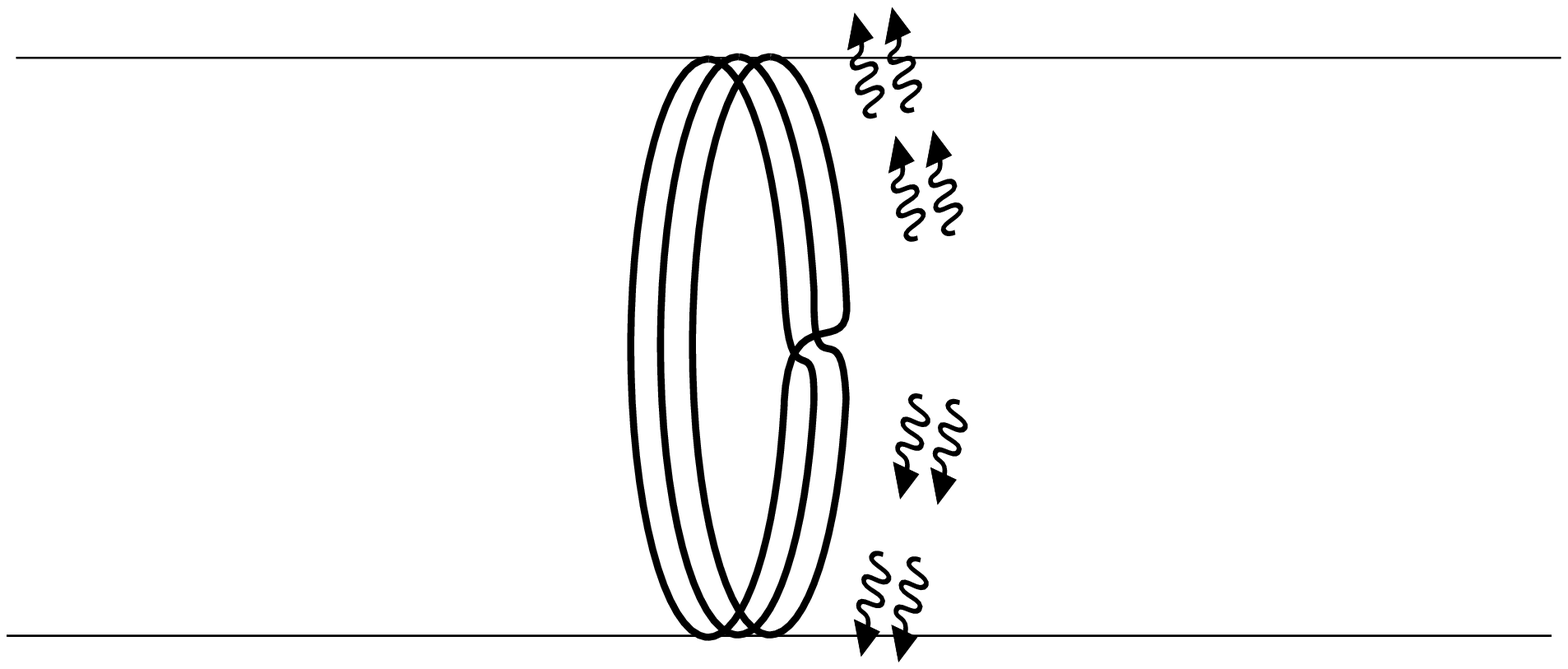} \\
    \vspace{.2truecm}
    \end{center}
    \hspace{3.7truecm} (a)\hspace{6.4truecm}(b)
   \caption{The NS vacuum is shown in (a) and the chiral primary state is shown in (b).}
   \label{fig:NSVaccumAndChiralPrimary}
\end{figure}

Let us start with the vacuum in the NS sector. Apply the operator (\ref{operator}).  This twist operator has linked together $k$ copies of the $c=6$ CFT. We call these linked copies a `component string'. The total central charge of this component string is
\be
c_{cs}=6k
\ee
Since the NS vacuum has $h=\bar h=j=\bar j=0$, we have obtained a state in the NS sector with
\be
\tilde \sigma_k^{--}|0\rangle_{NS}: ~~h=\bar h={k-1\over 2},~~~ j= \bar j=-{k-1\over 2}
\label{operatorp}
\ee
The NS vacuum and the state obtained after acting on the NS vacuum with operator \bref{operator} are shown in figure \ref{fig:NSVaccumAndChiralPrimary}

\subsubsection{ Spectral flow}

The field theory on the D1-D5 branes system is in the R sector.
This follows from the fact that the branes are solitons of
the gravity theory, and the fermions on the branes are induced
from fermions on the bulk. The latter are periodic around the
$S^1$; choosing antiperiodic boundary conditions would give a
nonvanishing vacuum energy and disallow the flat space solution
that we have assumed at infinity.

The NS sector
states can be mapped to R sector states by `spectral flow'
\cite{spectral}. This is a general symmetry of  under which the conformal dimensions and
charges change as
\bea
h'&=&h+\alpha j + \alpha^2{c\over 24}\\
j'&=&j+\alpha{c\over 12}
\label{qone}
\eea
Setting $\alpha=1$ gives the flow from the NS sector to the R
sector, and we can see that under this flow anti-chiral primaries
of the NS sector
(which have $h=-j$) map to Ramond ground states with $h={c\over
24}$.

If we set $\alpha=2$ in (\ref{qone}) then we return to the NS
sector, and setting $\alpha=3$ brings us again to the R sector.
More generally, the choice
\be
\alpha=(2n+1)\,, ~~~ n\in\mathbb{Z}
\label{qtwo}
\ee
brings us to the R sector.

\subsubsection{Spectral flow of the component string}

\begin{figure}[htbp] 
 \begin{center}\hspace{-1truecm}
  \includegraphics[width=2in]{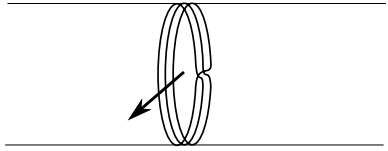} 
    \vspace{.2truecm}
    \end{center}
   \caption{The ground state in the twisted Ramond sector.}
   \label{fig:RossVacuum}
\end{figure}

Let us take the NS sector state (\ref{operatorp}) and spectral flow by $\alpha=1$ units to the R sector. This R sector state has
\be
h=\bar h={k\over 4}, ~~~j=\bar j={1\over 2}
\label{compareq}
\ee
This state has the lowest energy for its central charge, and thus carries no bosonic or fermionic excitations. We note that it does carry a charge of ${1\over 2}$ in each of the left and right sectors. We call this charge the `base spin' of the state. Once we add excitations, more charge will arise from the fermions on the component string, but this base spin will play a very important role. This state is shown in figure \ref{fig:RossVacuum}.

Now let us spectral flow by a further amount $\alpha=2n$. This will keep us in the R sector, but add excitations to the left and right sectors and make the component string non-extremal. Now we get
\be
h=\bar h= k (n^2 + \f{n}{k} + \f{1}{4}), ~~~j=\bar j= nk + \h
\label{cstring}
\ee
Comparing to (\ref{compareq}) we see that the extra spectral flow by $2n$ units has added an excitation with
\be
\Delta h=\Delta \bar h= k (n^2 + \f{n}{k})
\label{excess}
\ee
Our final state will consist of many component strings of this type.

\subsubsection{The states we consider}

The full CFT has $N$ copies of the $c=6$ CFT. The component string described above links together $k$ copies of the CFT. Let us consider the state of the CFT where all copies of the CFT are linked into such multiwound component strings of winding number $k$. Thus there will be ${N\over k}$ component strings, and we will have an R sector state
\be
[(\sigma^{--}_k)^{N\over k}|0\rangle_{NS}]_{\alpha_L=\alpha_R=(2n+1)}\equiv |\Psi^{--}(k,n)\rangle
\label{two}
\ee
with
\be
h=\bar h= N\Bigl(n^2+{n\over k}+{1\over 4}\Bigr), ~~j=\bar j={N\over
2}\Bigl(2n+{1\over k}\Bigr)
\label{twop}
\ee

Let us now identify the gravity solutions which are dual to these CFT states.

\bigskip

(i) This  CFT state  has no momentum since we have chosen equal spectral flow for the left and right movers: 
\be
n_p=h-\bar h=0
\ee
So the gravity solution will also be one with no momentum $n_p$. 

(ii) Recall that the angular directions of the noncompact space have a rotation group $so(4)\approx su(2)_L\times su(2)_R$. The $so(4)$ representations are characterized by azimuthal quantum numbers $J_\psi, J_\phi$. The $su(2)$ factors have azimuthal quantum numbers $j, \bar j$. The relation between these descriptions is \cite{cm1}
\bea
J^{cft}_\psi &=& -j -\bar j = -N( 2n+\f{1}{k}) \nn
J^{cft}_\phi &=& j - \bar j =0 \label{Eqn:JIntCFT}
\eea
We can now equate these angular momenta to the angular momenta of the geometries (\ref{Eqn:metric}) (eq. (\ref{Eqn:JInt})). From this we conclude that
\be
m=2nk+1 \label{Eqn:MandNK}
\ee

(iii) With these choices, we find that the energy agrees between the CFT state and the gravity solution. First consider {\it extremal} states. 
In the CFT, if all component strings were in the state (\ref{compareq}) then we would have an extremal state of the D1-D5 system since each component string will have the lowest energy for its central charge. In the gravity description, this would correspond to an extremal geometry, which has the minimum mass for its charge.

Now let us look at the energy {\it above} extremality. In the CFT state we have, using (\ref{excess})
\be
E- E_{ext} = \f{1}{R} {N\over k} (\Delta h+\Delta \bar h) = \f{2 N}{R} (n^2 + \f{n}{k}) \label{Eqn:DeltaMCFT}
\ee
We should compare this to $\Delta M_{ADM}$ (eq. (\ref{Eqn:DeltaM})) which gives the energy above extremality for the {\it gravity} solution. Using  (\ref{Eqn:MandNK}), we see that we get agreement.

\subsubsection{Fermionic excitations of the state}

Let us now give a  more explicit  description of the states (\ref{twop}). 

 Since all component strings are the same, let us restrict our attention to one of the component strings. This  string is  wound $k$ times around a circle of length $2 \pi R$. There are $\f{N}{k}$ such component strings. There will be no bosonic excitations; the excitation energy will all be carried by fermions alone. Since we are in the R sector, the fermions were periodic under $\sigma\r \sigma+2\pi R$ before application of the twist operator, and are therefore periodic under  $\sigma\r \sigma+2\pi k R$ on the multiwound component string.  Thus the fermions have energies  in multiples of $\f{1}{kR}$. Since both the left and the right sectors are same in our problem,  we look at just the left sector. 
 
\begin{figure}[htbp] 
 \begin{center}\hspace{-1truecm}
  \includegraphics[width=2in]{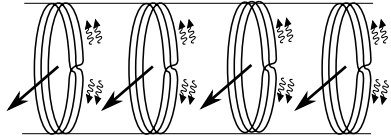} 
    \vspace{.2truecm}
    \end{center}
   \caption{The initial state of the CFT; each component string has a certain winding number and left and right fermionic excitations.}
   \label{fig:FullRoss}
\end{figure}

 We place one fermion of type $\psi$ with spin $j=\h$ in the lowest allowed energy level $\f{1}{kR}$, the next in the second level $\f{2}{kR}$ and so on till we occupy the energy level ${n\over R}$. These fermions have a total  energy
\be
E = \f{1}{kR}[ 1+ 2+ \dots + nk] = \f{n(nk+1)}{2R} 
\ee
and a spin
\be
j=\f{kn}{2}
\ee
The levels for the fermion $\tilde \psi$ are filled up in the same manner. For each component string we get from these fermions  an energy and charge (for the left sector)
\be
E_{cs}=\f{n(nk+1)}{R}, \qquad j_{cs}=kn
\ee
Recall that each component string had a base spin ${1\over 2}$ (eq. (\ref{compareq})) which we must add to the charge. We have  $\f{N}{k}$ identical component strings.   We then find for the full CFT state
\be
E^{cft}_L=E^{cft}_R=\f{N}{R}(n^2+ \f{n}{k}), \qquad j^{cft}_L=j^{cft}_R= \f{N}{2}(2n + \f{1}{k})
\ee
which agrees with (\ref{twop}).
This state is shown in figure \ref{fig:FullRoss}.

To summarize, we have given an explicit representation of our states in terms of excited component strings.  There are  $\f{N}{k}$ component strings, each with winding number $k$. We have two species of fermions filling up a fermi sea with no holes. Both the left and the right sectors have the same structure.  The states carry no bosonic excitations.

\subsubsection{The twist operator}

We now wish to examine the emission of a supergravity quantum from the D1D5 system in  the CFT description. 
The details of this process were listed in \cite{cm1}. The main change for our present case is the fact that the component strings in the initial state have winding number $k$ instead of winding number unity. 

The emission in the CFT is described by a vertex operator. We will consider the emission of scalars with angular momentum $l$. The rotation group is $so(4)\approx su(2)_L\times su(2)_R$, and the emitted quantum is in the representation 
\be
(\f{l}{2},\f{l}{2})
\ee
of $su(2)_L \times su(2)_R$. The vertex operator for this process has a twist operator of order $l+1$, so that $l+1$ copies of the CFT get joined into one copy at the point of insertion of the emission vertex. 

In our present case the component strings before emission already have a winding number $k$, so $k$ copies of the CFT are linked together in each component string. The $l+1$ copies of the CFT involved in the emission vertex can come from $l+1$ {\it different} component strings, or we might have more than one copy coming from the {\it same} component string. We are in the limit where $k\ll N$, where $N=n_1n_5$ is the total number of copies of the CFT. In this case we can assume that the $l+1$ copies come from $l=1$ {\it different} component strings; the probability for two copies to come from the same component string is down by a factor ${k\over N}$.

\begin{figure}[htbp] 
 \begin{center}\hspace{-1truecm}
  \includegraphics[width=2in]{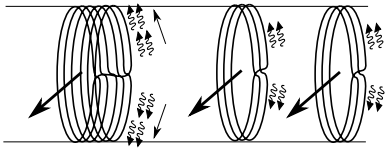} 
    \vspace{.2truecm}
    \end{center}
   \caption{The state of the CFT after one `final state' component string has been produced.}
   \label{fig:RossAfterEmission}
\end{figure}

In this situation the twist created by the emission vertex leads to the formation of one component string with winding number $k(l+1)$. 
Now excitation energies  on this component string can come in fractional units ${1\over k(l+1)R}$. Recall that figure (\ref{fig:FullRoss}) showed an initial state of the CFT {\it before} any emission. In figure \ref{fig:RossAfterEmission} we depict the state {\it after} one quantum has been emitted and a component string with winding $k(l+1)$ has been produced. The figure depicts the case $k=3, l=1$, so that the initial component strings have winding number $k=3$ and the number of component strings joined together is $l+1=2$.

\subsection{Obtaining $\omega_R$ from the CFT}

We will start with the initial state of the CFT (\ref{two}). We will construct  a set of simple final states for the CFT, and observe that the change of energy and charge agrees with the spectrum of emitted quanta in the {\it gravity} computation of emission.

\subsubsection{The energy spectrum to be reproduced}

Our main goal is to show how the winding number $k$ of the initial state component strings enters the computations. Thus we will take a simple set of emission possibilities, rather than the full set studied for the case $k=1$ in \cite{cm1}. The emitted quantum is characterized by angular quantum numbers $m_\phi, m_\psi$, a momentum $\lambda$ along $S^1$, and an energy $\omega$. 
We will look at the emissions with
\be
m_\psi=-l, ~~m_\phi=0, ~~\lambda=0, ~~\omega=\omega_{max}
\label{first}
\ee
where $\omega_{max}$ is the largest possible emission energy for the given quantum numbers $m_\psi, m_\phi, \lambda$.
The computations can be easily extended to other values of these parameters by following the computations in \cite{cm1}. 

With (\ref{first}), the frequencies of the emitted quanta in the gravity computation are
\be
\omega_R = \f{1}{k R} (-l-2-m_\psi m)=\f{2}{R}(nl-\f{1}{k})
\label{omegaqq}
\ee
where we have used $m=2nk+1$ (eq. \bref{Eqn:MandNK}).

\subsubsection{Properties of the emitted quantum in the CFT description}

In the CFT the energy and charge of the emitted quantum will show up as a decrease in energy and charge of the D1D5 state:
\be
\omega_{scalar}={1\over R} [ (h_i-h_f)+(\bar h_i-\bar h_f)]\equiv {1\over R} (\Delta h+\Delta\bar h)
\ee
Note that $\lambda$ is the momentum along the $S^1$, which is the spatial direction of the CFT. Since we have taken $\lambda=0$, the left and right contributions to the emitted scalar will be equal ($\Delta h=\bar \Delta h$). From  (\ref{omegaqq}) we see that we need
\be
\Delta h=\Delta\bar h=nl-{1\over k}
\label{deltahqq}
\ee
The angular momenta of the emitted scalar are given in terms of the change in charges of the CFT state:
\be
m_\psi=-\Delta j-\Delta \bar j, ~~~m_\phi=\Delta j-\Delta \bar j
\ee
Since we have taken $m_\psi=-l, m_\phi$ =0, we have 
\be
\Delta j=\Delta \bar j={l\over 2}
\label{deltajqq}
\ee
Our goal will be to see if the allowed final states of the CFT give
(\ref{deltahqq}),(\ref{deltajqq}).

\subsubsection{Computing the CFT quantum numbers}

The initial state is symmetric between the left and right movers, and our choice of emitted quantum also has this symmetry. Thus the final state will also have equal left and right quantum numbers. Thus in what follows we write only the left quantum numbers at each step.

\bigskip

{\it The initial state:} \quad Each component string in the initial state has $h, j$ given by (\ref{cstring}). We need to consider $l+1$ such component strings in our process, and can ignore the other component strings.  For these $l+1$ component strings, we have
\be
j_i=(l+1)( nk + \h)
\ee
\be
h_i = (l+1) \f{nk(nk+1)}{k} + k  \f{l+1}{4} 
\ee

\bigskip

{\it The final state:} \quad We will postulate a choice of final state, and then observe that it has the required quantum numbers to account for the quantum numbers of the emitted quantum. Let the final state be described as follows:

(i) The $l+1$ component strings, each with winding number $k$, have been joined into one long string with winding number $(l+1)k$.

(ii) The fermions of type $\psi$ occupy the energy levels
\be
 {1\over (l+1)kR},~~ {2\over (l+1)kR}, ~~\dots ~~{nk(l+1)\over (l+1)kR}
 \label{levels}
 \ee
 Thus there are $nk(l+1)$ fermions of this type. Each contributes a charge ${1\over 2}$. Similarly we place the fermions of type $\tilde\psi$ in levels (\ref{levels}), again getting a charge ${1\over 2}$ from each fermion. Finally,, we have the base spin ${1\over 2}$ for the single long component string. Thus the charge of the final state is
 \be
 j_f={1\over 2}[(l+1)nk + (l+1)nk]+\h=j_f=(l+1)nk + \h
\ee

(iii) In \cite{cm1} it was argued that the final state needs one bosonic excitation $\p X^i$ for the left movers  and one bosonic excitation $\bar  \p X^j$ for the right movers. This is because the emitted scalar in the full string theory arises from a graviton $h_{ij}$ with both indices on the torus $T^4$, and the vertex operator for emission creates excitations with quantum numbers $i, j$ in the final state. In the case of \cite{cm1} the energy of these bosonic excitations was ${1\over R}$ each; thus it was the lowest allowed energy on the component strings before the twisting process makes the string longer. By analogy, now our component strings start off with winding number $k$, so we will let the dimension of the left and right bosonic excitations be
\be
h_b=\bar h_b={1\over k}
\label{hboson}
\ee

Let us now write down the dimension $h_f$ of the final state. The Ramond ground state of the component string has a base dimension ${c\over 24}$, where $c$ is the central charge of all the copies of the CFT in our state. There are $k(l+1)$ linked copies, each with $c=6$, so this contribution is
\be
h=  k \f{l+1}{4} 
\ee
The fermion of each type in levels (\ref{levels}) gives
\be
h=\sum_{s=1}^{nk(l+1)}{s\over (l+1)kR}= {1\over 2}\f{1}{k(l+1)}(nk (l+1) (nk(l+1)+1) 
\ee
The bosons give
\be
h={1\over k}
\ee
Thus the dimension of the final state is
\be
h_f= \f{1}{k(l+1)}(nk (l+1) (nk(l+1)+1) + k \f{l+1}{4} + {1\over k}
\ee

\bigskip

{\it Quantum numbers of the emitted scalar:} \quad With the above postulate for the structure of the final state, we find for the emitted quantum
\bea
j_{scalar} &=& j_i - j_f = \f{l}{2} \nn
h_{scalar} &=& h_i -h_f = nl - {1\over k}
\eea
We now observe that these quantum numbers agree exactly with the requirements (\ref{deltajqq}),(\ref{deltahqq})
arrived at from the {\it gravity} computation. Thus we have generalized the result of \cite{cm1} on the emission frequencies to the case $k\ne 1$; i.e., the emission frequencies from the CFT state agree with the instability frequencies of the corresponding non-extremal geometry.

\subsection{Obtaining $\omega_I$ from the CFT}

In the previous subsection we saw how the {\it spectrum} of emission obtained from gravity is reproduced from the CFT. In this section we will show how to reproduce the emission {\it rate} from the CFT. This calculation was performed in  detail in \cite{cm1} for the case $k=1$. 
Here we will list the changes involved when the computation is done for  general $k$.

To compute $\omega_I$ from the CFT we consider the interaction where excitations on the component strings collide and produce a supergravity quantum. The emission rate  depends on an emission {\it vertex} which depends on the dynamics of the CFT, and on the occupation numbers of various excitations on the component strings. The emission vertex can be found from the CFT, but so far its overall normalization has not been computed, except for the case $l=0$. When we choose thermal distributions for the excitations, the CFT emission rate is found to agree with the semiclassical {\it Hawking} radiation from black holes, up to the overall undetermined constant \cite{radiation, angular}. Our strategy in \cite{cm1} was to fix this overall normalization constant $V(l)$ by requiring the CFT emission to agree with the Hawking radiation rate when thermal distributions are chosen for the CFT excitations. Then we replace the thermal distributions by the distributions that we actually have for our special microstate, and recompute the emission. It was found that the result matched exactly with the instability radiation found for the gravity description of the microstate. Here we wish to look for a similar agreement for general $k$.

We will now recall the steps of the computation in \cite{cm1}, and point out where the factors of $k$ appear when we generalize to $k\ne 1$.

\begin{itemize}
\item

For the case $k=1$, The  amplitude  ${\cal R}$  for emission from the CFT state was found in \cite{cm1}   (eq.(4.34))
\be
\mathcal R_{k=1} = V(l) \f{1}{4 \pi} \sqrt{\f{\pi}{r_{max}}} \f{1}{(2\pi R)^{l+\h}} \omega^{l+1} \sqrt{\omega_1 \bar \omega_1} \delta_{\sum n_i-\sum \bar n_i-\lambda,0} \prod \mathcal D \label{Eqn:RfromCM1}
\ee
Here $\omega_1,\bar \omega_1$ are the left and right energies of the bosonic excitations on a component string, $n_i,\bar n_i$ are the levels of the bosonic and fermionic excitations on the string and the $\mathcal D$ give occupation numbers for excitations on the component strings. The radius $r_{max}$ comes from regularizing the box size in which the scalar is emitted and cancels out at the end. In the case where the component strings have winding number $k$, we get
\be
\mathcal R_{k} = V_k(l) \f{1}{4 \pi} \sqrt{\f{\pi}{r_{max}}} \f{1}{(2\pi k R)^{l}} \f{1}{\sqrt{ 2 \pi R}} \omega^{l+1} \sqrt{\omega_1 \bar \omega_1} \delta_{\sum n_i-\sum \bar n_i-\lambda,0} \prod \mathcal D \label{Eqn:RModified}
\ee
To obtain this answer note that the  two bosonic and $2l$ fermionic excitations now live in a `box' of length $2 \pi k R$ instead of $2\pi R$, and the field operators thus have factors ${1\over \sqrt{2\pi k R}}$  instead of ${1\over \sqrt{2\pi R}}$ multiplying the creation and annihilation operators.  Integrating the emission vertex over the `box' now gives $2\pi k R$ instead of $2\pi R$. Putting these two effects together gives the power ${1\over k^l}$ in (\ref{Eqn:RModified}).

\item
In \cite{cm1} (eq.(7.17)) the emission amplitude \bref{Eqn:RfromCM1} was used to find the emission {\it rate} 
\be
\Gamma(\omega)_{k=1} = \f{|V(l)|^2}{8 \pi} \f{\omega^{2l+2} }{(2\pi R)^{2l+1}} \sum_{state} \omega_1 \bar \omega_1 \delta_{\sum \omega_i, \f{\omega}{2}} \delta_{\sum \bar \omega_i, \f{\omega}{2}}   \prod \mathcal D^2 \label{Eqn:RatefromCM1}
\ee
(The parameter $\lambda$ in \bref{Eqn:RfromCM1} has been chosen to be zero for simplicity.)
For general $k$ we get
\be
\Gamma(\omega)_{k}  = \f{|V_k(l)|^2}{8 \pi} \f{\omega^{2l+2}}{(2\pi k R)^{2l}} \f{1}{2\pi R}\sum_{state} \omega_1 \bar \omega_1\delta_{\sum \omega_i, \f{\omega}{2}} \delta_{\sum \bar \omega_i, \f{\omega}{2}}  \prod \mathcal D^2 \label{Eqn:Rate}
\ee
where the factor of $\f{1}{k^{2l}}$ comes from the step where the amplitude \bref{Eqn:RModified} is squared to get the emission probability and thus the rate of emission.

\item
In \cite{cm1} the expression \bref{Eqn:RatefromCM1} is then evaluated for the case where the energies of excitations on the CFT are very high compared to the energy gap on the component strings. In this limit the sum over the initial states can be replaced by an integral. In  \cite{cm1} (eq.(7.18)) we had 
\be
\sum_{\omega_i} \to R \int d\omega_i \label{Eqn:sumToIntegralCM1}
\ee
because the length of the component string was $2 \pi R$. This gave (eq.(7.19) in \cite{cm1})
\be
\Gamma(\omega)_{k=1}=  \f{|V(l)|^2}{8 \pi} \f{\omega^{2l+2} \omega_1 \bar \omega_1}{(2\pi )^{2l+1} R}  \left[ \int_{-\infty}^\infty \prod_{i=1}^{l+1} d \omega_i d \bar \omega_i~ \omega_1 \bar \omega_1 \delta(\f{\omega}{2}- \sum \omega_i) \delta(\f{\omega}{2}- \sum \bar \omega_i) \prod \mathcal D^2 \right]
\ee
In the present case  \bref{Eqn:sumToIntegralCM1} becomes
\be
\sum_{\omega_i} \to k R \int d\omega_i \label{Eqn:sumToIntegral}
\ee
Converting the Kronecker delta functions to Dirac delta functions brings in a factor of $k^{-1}$ for each delta function. We get
\be
\Gamma(\omega)_{k}=  \f{|V(l)|^2}{8 \pi} \f{\omega^{2l+2} \omega_1 \bar \omega_1}{(2\pi )^{2l+1} R}  \left[ \int_{-\infty}^\infty \prod_{i=1}^{l+1} d \omega_i d \bar \omega_i~ \omega_1 \bar \omega_1 \delta(\f{\omega}{2}- \sum \omega_i) \delta(\f{\omega}{2}- \sum \bar \omega_i) \prod \mathcal D^2 \right]
\ee
Note that at this step all powers of $k$ have cancelled out.

\item
Consider emission for the case where the excitations have the thermal distributions appropriate for the {\it generic} black hole state. The  $\mathcal D^2$ describe standard bose and fermi distributions and do not change for the case of general $k$. (Thus eqs. (7.30),(7.10) in  \cite{cm1} remain unchanged.) We do find a change when we replace the sum by an integral for the energies of the emitted quantum. Eq. (7.32) in \cite{cm1} 
\be
\sum_{\omega} \to  \int \f{R}{2} d \omega
\ee  
changes to 
\be
\sum_{\omega} \to  \int  k \f{R}{2} d \omega
\ee
This changes the expression for emission into the energy interval $d\omega$. For the case $k=1$ this was  eq.(7.33) in  \cite{cm1}
 \bea
d \Gamma_{l_{k=1}}&=&\f{ |V(l)|^2 \omega^{2l+2} ~d\omega }{8 \pi (2 \pi )^{2l+1}} \f{1}{[(l+1)!]^2 2^{2l+3}}\nn
&&~~~~~~\times [\omega^2 +(2\pi T_L)^2 1^2][\omega^2 +(2\pi T_L)^2 3^2]\dots [\omega^2 +(2\pi T_L)^2 l^2]\nn
&&~~~~~~\times [\omega^2 +(2\pi T_R)^2 1^2][\omega^2 +(2\pi T_R)^2 3^2]\dots [\omega^2 +(2\pi T_R)^2 l^2]\nn
&&~~~~~~\times {1\over (e^{\omega\over 2 T_L}+1)(e^{\omega\over 2 T_L}+1)} \label{Eqn:dGammaLCM1}
\eea
For general $k$ this changes to
 \bea
d \Gamma_{l_k}&=&\f{k |V(l)|^2 \omega^{2l+2} ~d\omega }{8 \pi (2 \pi )^{2l+1}} \f{1}{[(l+1)!]^2 2^{2l+3}}\nn
&&~~~~~~\times [\omega^2 +(2\pi T_L)^2 1^2][\omega^2 +(2\pi T_L)^2 3^2]\dots [\omega^2 +(2\pi T_L)^2 l^2]\nn
&&~~~~~~\times [\omega^2 +(2\pi T_R)^2 1^2][\omega^2 +(2\pi T_R)^2 3^2]\dots [\omega^2 +(2\pi T_R)^2 l^2]\nn
&&~~~~~~\times {1\over (e^{\omega\over 2 T_L}+1)(e^{\omega\over 2 T_L}+1)} \label{Eqn:dGamma}
\eea
The above expressions are for odd $l$; a similar analysis holds for even $l$. 

\item
Next we recall the semiclassical Hawking emission rate (eq.(7.6) in \cite{cm1})
\bea
d \Gamma_l&=&{\pi\over 8}{(Q_1Q_5)^{l+1}\over 2^{4l} [l!(l+1)!]^2}
\omega^{2l+2}d\omega\nn
&&~~~~~~\times [\omega^2 +(2\pi T_L)^2 1^2][\omega^2 +(2\pi T_L)^2 3^2]\dots [\omega^2 +(2\pi T_L)^2 l^2]\nn
&&~~~~~~\times [\omega^2 +(2\pi T_R)^2 1^2][\omega^2 +(2\pi T_R)^2 3^2]\dots [\omega^2 +(2\pi T_R)^2 l^2]\nn
&&~~~~~~\times {1\over (e^{\omega\over 2 T_L}+1)(e^{\omega\over 2 T_L}+1)} \label{Eqn:GravGammaOddl}
\eea
For $k=1$ we equate  \bref{Eqn:dGammaLCM1} to  \bref{Eqn:GravGammaOddl} and get (eq. (7.35) in \cite{cm1})
\be
|V(l)|^2 = \f{16 \pi^{2l+3}}{[l!]^2} (Q_1 Q_5)^{l+1}
\ee

Now consider the situation for general $k$.  In modelling the black hole by a CFT state, we have to use a thermal distribution of excitations, which means that we have to assume that the energy of the typical quantum is much larger than the energy gap on the component string. Given this limit, what length we choose for the component strings is irrelevant. While we had used $k=1$ for each component string, in \cite{cm1}, now we will let the winding number  be $k$. This will alter the details of the emission vertex, and we have seen  above  how factors of $k$ arose in the various steps needed in computing the emission. The overall normalization of the vertex will also have to be recomputed by again requiring agreement with the semiclassical Hawking emission rate. 
Equating \bref{Eqn:dGamma} with  \bref{Eqn:GravGammaOddl} we get
\be
|V_k(l)|^2 =\f{1}{k} \f{16 \pi^{2l+3}}{[l!]^2} (Q_1 Q_5)^{l+1} \label{Eqn:VertexNormalization}
\ee
If we have to study emission from any state where the component strings have winding number $k$, we must use the above normalization of the vertex.

\end{itemize}

\bigskip

Using \bref{Eqn:VertexNormalization} in \bref{Eqn:Rate} we get
\be
\Gamma(\omega)_{k}  =   \f{4 \pi k R}{[l!]^2} \left( \omega^2 \f{Q_1Q_5}{4(k R)^2} \right)^{l+1} \sum_{state} \omega_1 \bar \omega_1 \delta_{\sum \omega_i, \f{\omega}{2}} \delta_{\sum \bar \omega_i, \f{\omega}{2}}  \prod \mathcal D^2 
\ee
This is the expression we were seeking; we have found the factors of $k$ that appear in the emission rate when the component strings have winding number $k$ each.

Now let us use this expression to get the emission from our actual microstate.
 For the energies of the bosons we have
from \bref{hboson}
\be
\omega_1  = \bar \omega_1 = \f{1}{k R}
\ee
For our special microstate all energy levels that are occupied are occupied with probability unity. This sets $\prod{\mathcal D}^2=1$ \cite{cm1}. We thus find
\be
\Gamma = \f{1}{kR} \f{4 \pi}{(l!)^2} \left( \omega^2 \f{Q_1 Q_5}{4 (kR)^2} \right)^{l+1}
\ee

 $\Gamma$  gives  the rate of spontaneous emission from the CFT state. As explained in \cite{cm1}, after $N_1$ final state component strings have been created, the probability of creating the next one gets a bose enhancement factor $N_1$. This leads to an exponential growth of the emission rate $\sim e^{\omega_I t}$, with
 \be
 \omega_I={\Gamma\over 2}
  \ee
Thus we finally obtain
\be
\omega_I =  \f{1}{kR} \f{2 \pi}{(l!)^2} \left( \omega^2 \f{Q_1 Q_5}{4 k^2 R^2} \right)^{l+1}
\ee
To compare to the gravity result for $\omega_I$ in \bref{Eqn:GravOmega}, let us recall that $\gamma={1\over k}$, and $\kappa^2=\omega^2{Q_1Q_5\over R^2}$. Then we observe that we get full agreement with \bref{Eqn:GravOmega}. Thus we have found that the emission rate from the CFT state agrees with the behavior of the instability radiation for the dual {\it gravity} solution.

\section{Emission from generic stationary geometries}
\setcounter{equation}{0}

The general microstate geometry is complicated, and we will not be able to solve the wave equation in it in closed form. But we still expect that the general idea of ergoregion emission should be applicable to a large class of non-extremal geometries. In this section we write down a general formalism which will give the rate of ergoregion emission {\it if} we are given the solutions to the wave equation in the `cap' region of the geometry.

The essential idea is to break up the geometry into three parts:

\bigskip

(i) The flat space region at large $r$.

(ii) A region that behaves approximately like $AdS_3\times S^3$. This region starts at the `neck' where the geometry departs from being flat, and ends before reaching the `cap'.  

(iii) The `cap region, where the details of the particular microstate manifest themselves in the geometry.

\begin{figure}[htbp] 
 \begin{center}\hspace{-1truecm}
  \includegraphics[width=2in]{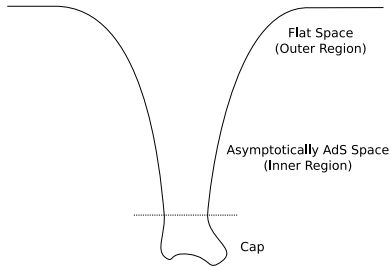} 
    \vspace{.2truecm}
    \end{center}
   \caption{A generic stationary geometry (in the large $R$ limit).}
   \label{fig:C&D}
\end{figure}

\bigskip
This geometry is shown schematically in figure \ref{fig:C&D}.
To study ergoregion emission from the microstate we need to solve the scalar wave equation in the geometry. For any choice of frequency $\omega$ there will be a solution to the wave equation which is everywhere regular in the `cap'. In the region (ii) the solutions of the wave equation will be simple since the geometry is just $AdS_3\times S^3$. For a given angular harmonic and a given $\omega$, there are two solutions of the second order scalar wave equation. Let us denote the coefficients of these two solutions by $F_1, F_2$. Requiring regularity of the solution in the `cap' will determine the ratio $\left({F_1\over F_2}\right )(\omega)$. We do not solve the wave equation in the cap, but assume that the net effect of the cap geometry is given to us as this ratio $\left({F_1\over F_2}\right )(\omega)$.

We can now match this solution in region (ii) to the solution in the flat space region (i). We put outgoing boundary conditions in region (i), as required for computing the ergoregion instability. Generalizing the method of \cite{myers}, we get an expression for the instability spectrum in terms of the ratio $\left({F_1\over F_2}\right )(\omega)$.

Let us now carry out these steps.

\subsection{The geometry}

We will work as before in the large $R$ limit (\ref{Def:epsilon}). The flat space geometry at large $r$ changes over to a nontrivial metric at
\be
\rho = \f{r R}{\sqrt{Q_1 Q_5}} \approx \f{R}{(Q_1 Q_5)^\f{1}{4}}
\ee 
The region (ii) will be given by
\be
1\ll \rho \ll \f{R}{(Q_1 Q_5)^\f{1}{4}}
\ee
In this region the geometry is locally $AdS_3\times S^3$. The effect of the cap is felt only through the angular momentum of the geometry. The metric for region (ii) will thus have the approximate form
\be
ds^2 \approx \rho^2(- d\tau^2 + d \varphi^2) + \f{d \rho^2}{\rho^2} + d \theta^2 + \cos^2 \theta ( d \psi + \alpha d \tau + \beta d \varphi)^2 + \sin^2 \theta (d \phi + \beta d \tau  + \alpha d \varphi)^2
\label{regionii}
\ee
where the parameters $\alpha, \beta$ depend on the angular momenta. The geometries of \cite{ross} have such a form in their region of type (ii), but now we are extending our analysis to geometries with more general cap structures.

\subsection{The wave equation}

We have to solve the wave equation $\square\Phi=0$. We will solve it in the region (ii), and in the flat space region (i), and match the two solution to find the instability frequencies.

\subsubsection{The region (ii)}

In the region (ii) with metric (\ref{regionii}) we can write the ansatz
\be
\Phi = e^{-i \omega R \tau + i \lambda R \varphi + i m_\psi  \psi + i m_\phi \phi} \chi(\theta) h (\rho)
\ee
We thus get
\be
  \f{1}{\rho} \partial_\rho(\rho^3 \partial_\rho h) + {\eta^2 \over \rho^2}+(1-\nu^2) h~=~0
\ee
Using the variable $x = \rho^2$ we get
\be
4 \partial_x(x^2 \partial_x h) + \f{\eta^2}{x} h+ (1 - \nu^2)h=0
\label{waveii}
\ee
where $\eta^2 = (\omega R+ \beta m_\phi + \alpha m_\psi)^2 -(\lambda R + \beta m_\psi + \alpha m_\phi)^2$. We also have
\be
\nu \approx l+1+O(\epsilon^2)
\ee
where $\epsilon$ defined in \bref{Def:epsilon} goes to zero as we take  $R$ to infinity. To understand the origin of the correction $\epsilon$,   recall that $\nu$ arises from the angular laplacian. The metric in angular directions departs from the exact sphere metric obtained for the angular directions from  (\ref{regionii}) because the metric of region (ii)  changes to the flat metric at large $r$. The value of $\epsilon$ was computed in \cite{hottube} for the geometries studied there, but for us the value of $\epsilon$ will not be important. We will take $R\rightarrow\infty$ at the end, so $\epsilon$ will go to zero. We keep $\epsilon$ in our computation as a regulator, since otherwise we will encounter poles in certain $\Gamma$ functions; the final expressions for the instability frequencies will have no such poles, and we will set $\epsilon=0$ at the end.

In region (ii) we had $\rho\gg1$, which gives $x=\rho^2 \gg 1$. The wave equation (\ref{waveii}) thus becomes
\be
4 \partial_x(x^2 \partial_x h) + (1 - \nu^2)h=0
\ee
with the
 general solution 
\be
h_{in} = \f{1}{\sqrt{x}} \left[ F_1 x^\f{\nu}{2} + F_2 x^{-\f{\nu}{2}} \right]
\label{middlesol}
\ee
with constants $F_1, F_2$.

\subsubsection{The region (i)}

The outer region (region (i)) has the flat metric
\be
ds^2 =- dt^2 + dr^2 + dy^2 + dr^2 ( d \theta^2 + \cos^2 \theta d\psi^2 + \sin^2 \theta d \phi^2)
\ee
Recall here that the variables $t, y$ in region (i) are related to the variables $\tau, \varphi$ used in region (ii) by the relation 
\be
\tau = \f{t}{R}, \qquad  \varphi = \f{y}{R}
\ee
The wave equation for region (i) is
\be
4 \partial_x(x^2 \partial_x h) +\kappa^2 x h+ (1 - \nu^2)h
\ee
where $\kappa^2 = (\omega^2 - \lambda^2) \f{Q_1 Q_5}{R^2}$. This has the solution
\be
h_{out} = \f{1}{\sqrt{x}} \left[  D_1 J_{\nu} ( \kappa \sqrt{x})+D_2 J_{-\nu} (\kappa \sqrt{x})  \right]
\label{innersol}
\ee

\subsubsection{Matching}

For small values $\kappa\sqrt{x}$ the solution (\ref{innersol}) is
\be
h_{out} = \f{1}{\sqrt{x}} \left[  D_1  \f{1}{\Gamma(1+\nu)} ( \f{\kappa \sqrt{x}}{2})^\nu+D_2 \f{1}{\Gamma(1-\nu)} (\f{\kappa \sqrt{x}}{2})^{-\nu}  \right]
\ee
Matching this to the solution (\ref{middlesol}) in region (ii) we get
\bea
F_1 &=& \f{D_1}{\Gamma(1+ \nu)}  \left( \f{\kappa}{2} \right)^{\nu} \nn
F_2 &=&  \f{D_2}{\Gamma(1-\nu)} \left( \f{\kappa}{2} \right)^{-\nu} 
\eea
which gives
\be
F \equiv \f{F_1}{F_2} = \f{D_1}{D_2} \f{\Gamma(1-\nu)}{\Gamma(1+ \nu)} \left( \f{\kappa}{2} \right)^{2 \nu}
\label{fff}
\ee

Now let us impose the boundary condition that we have purely outgoing waves at infinity \cite{myers}. From (\ref{innersol}) we get
\be
D_1 + D_2 e^{-i \pi \nu}=0
\ee
This gives
\be
F= - e^{- i \pi \nu}  \f{\Gamma(1-\nu)}{\Gamma(1+ \nu)} \left( \f{\kappa}{2} \right)^{2 \nu}
\label{match}
\ee

\subsection{Solving for the frequencies}

Recall that $\kappa \sim \omega$, and that for large $R$ we get small frequencies $\omega$. Thus the factor  $\left( \f{\kappa}{2} \right)^{2 \nu}$ appearing on the RHS of (\ref{match}) is very small.  Following the method of \cite{myers}, we say that to leading order in ${1\over R}$ the RHS of (\ref{match}) is {\it zero}. Then the relation (\ref{match}) will hold only for those frequencies $\omega_0$ which satisfy
\be
F(\omega_0) = 0
\label{cond}
\ee
From (\ref{fff}) we see that this means that $F_1(\omega_0)=0$, and from (\ref{middlesol}) we see that this means that the wavefunction in region (ii) has only a decaying part at large $r$. Thus the condition (\ref{cond}) just tells us that to leading order in ${1\over R}$ the solution chosen in the cap region must be such that it dies off towards the boundary of the $AdS$ region. This means that we would  get   a normalizable state if the geometrically was really asymptotically $AdS$ (instead of changing over to flat space at some large $r$). In other words,  the frequencies $\omega_0$ of the emitted quanta will be just the frequencies of the normal modes of the scalar field in the asymptotically $AdS$ geometry obtained for $R\r \infty$.

Note that $\omega_0$ is real.    Now we iterate (\ref{match}) to the next order, to find the {\it imaginary} part of the frequency $\omega$; this part gives the rate at which the instability {\it grows}. Generalizing the method of \cite{myers}, we look at values of $\omega$ near $\omega_0$ where $F(\omega)$ is not zero but equal to the small RHS of (\ref{match}). We get
\be
F(\omega_0 + \delta \omega)= F'(\omega_0) \delta \omega = - e^{- i \pi \nu}  \f{\Gamma(1-\nu)}{\Gamma(1+ \nu)} \left( \f{\kappa_0}{2} \right)^{2 \nu}
\label{final}
\ee
where $\kappa_0$ is $\kappa$ evaluated at $\omega_0$. 

We use the relation
\be
\Gamma(\nu) \Gamma(-\nu) = -\f{\pi}{ \nu \sin (\pi \nu)}
\ee
to find
\be
\delta\omega=-{e^{-i\pi\nu}\over F'(\omega_0)}{\Gamma(1-\nu)\over \Gamma(1+\nu)}
\left( \f{\kappa_0}{2} \right)^{2 \nu}
\ee
Taking the imaginary part, we get
\be
\omega_I=\delta \omega_I = \f{2 \pi}{\Gamma(\nu)^2} \left( \f{\kappa_0}{2} \right)^{2 \nu} \left( \f{1}{2 \nu F'(\omega_0)} \right) \label{Eqn:ImagOmega}
\ee
If we find that $\omega_I>0$, then we have an unstable mode, and $\omega_I$ gives the growth rate of the instability. (If $\omega_I<0$ the same computation gives the rate of emission of particles with energy $\omega_0$, if we had placed such quanta in the interior of the geometry \cite{cm2}.)

Thus we have expressed the growth rate of the instability in terms of the ratio $F(\omega)$. This function is to be found for any given microstate by  solving the scalar field equation in the `cap', and then the above formalism gives the ergoregion emission from the full geometry in the limit where $R$ is assumed large.

\section{Another family of smooth geometries with ergoregion}

We have seen that we can get emission from microstates by the process of ergoregion emission. It is therefore interesting to make more examples of geometries which do have an ergoregion. The geometries of \cite{ross} arise by considering the states with maximal angular momenta for given charges and mass. In this section we describe another family of geometries which have an ergoregion but somewhat lower rotation. We will proceed in the following steps:

\bigskip

(a) We will start a particular microstate family for the {\it extremal} 2-charge D1-D5 system. We take the large $R$ limit as before. We can now `chop off' the flat space part of the geometry at large $r$, getting an asymptotically $AdS$ geometry. 

(b) Next, we perform a spectral flow on both the left and right sectors of the microstate. In the CFT this adds energy to both the left and the right movers, and makes the state nonextremal. In the gravity description, the spectral flow is a {\it coordinate change} \cite{balmm}.
This coordinate change changes the time coordinate in particular, and an extremal solution changes to a nonextremal one.

(c) To get an asymptotically flat geometry for this nonextremal microstate, we should `re-attach' that flat part of the geometry. It is not clear how to construct the `neck' region where the $AdS$ space changes over to flat space. But we have seen that if $R$ is large, the details of the `neck' region are {\it not} relevant to the process of ergoregion emission. This is because the wavelength of the emitted quantum scales as $\sim R$, while the `neck' region extends over a distance of order $(Q_1Q_5)^{1\over 4}$. Thus if we are in the large $R$ limit \bref{Def:epsilon}
\be
 \f{\sqrt{Q_1Q_5}}{R^2} \ll  1 
\ee
then the emitted wave crosses over from the $AdS$ part of the geometry to the flat part without distortion. Thus we have to `match' solutions only between an $AdS$ region and a flat space region, and
this match gives the ergoregion emission. The important part of the computation was the change of time coordinate under spectral flow. The relation between the time coordinate inside and the time coordinate at infinity determines if there is an ergoregion, and the consequent instability and radiation. Thus all we have to do is locate the ergoregion using the correct time coordinate.

\subsection{The extremal geometry}
 
The extremal geometry we wish to use was constructed in \cite{lm4}.  General 2-charge extremal  geometries are constructed by starting with the 
NS1-P system, which is just a string carrying momentum. To get extremal geometries in the family of \cite{ross} the vibration profile of the NS1 would be a uniform helix. Now we will take instead the profile where the first half of the NS1 describes a helix in the clockwise direction, and the second half a helix in the anticlockwise direction. The angular momenta of this NS1 therefore totals to zero. We then perform S,T dualities to get the corresponding 2-charge extremal D1-D5 geometry. The resulting extremal  geometry is
\cite{lm4}
\be
ds^2 = \f{-f}{\sqrt{\tilde H_1 \tilde H_5}} (dt^2 -dy^2) + \sqrt{\tilde H_1 \tilde H_5} ( \f{dr^2}{r^2 + a^2} + d\theta^2) + \f{\sqrt{\tilde H_1 \tilde H_5}}{f}(( r^2+a^2) \cos^2 \theta d \psi^2 + r^2  \sin^2 \theta d \phi^2)
\ee
where
\be
f = r^2 + a^2 \sin^2 \theta, \qquad  \tilde H = f+ Q_i
\ee
and
\be
a = \gamma \f{\sqrt{Q_1 Q_5}}{R} 
\ee
This time we have 
\be
\gamma = \f{1}{2k}
\label{gamma2k}
\ee
 where $k$ gives the number of turns of the helix described by each half of the NS1 (in the NS1-P duality frame).

Let $r$ be small enough so that
 $f\ll Q_i$. This corresponds to chopping off the flat space region at large $r$. The metric takes the form
\be
ds^2 =\f{-f}{\sqrt{Q_1 Q_5}} (dt^2 -dy^2) + \sqrt{Q_1 Q_5} ( \f{dr^2}{r^2 + a^2} + d\theta^2) + \f{\sqrt{Q_1 Q_5}}{f}(( r^2+a^2) \cos^2 \theta d \psi^2 + r^2 \sin^2 \theta d \phi^2)
\ee
We work in the scaled coordinates better suited for this region
\be
\tau = \f{t}{R}, \qquad \varphi = \f{y}{R}, \qquad \rho = \f{r R}{\sqrt{Q_1 Q_5}} = \gamma \f{r}{a}
\ee
The metric is then
\bea
ds^2 &=& \sqrt{Q_1 Q_5} \Bigg[ - (\rho^2 + \gamma^2 \sin^2 \theta) ( d\tau^2 - d \varphi^2)+ \f{d \rho^2}{\rho^2 + \gamma^2} \nn
&&+ d \theta^2 + \f{\rho^2 + \gamma^2}{\rho^2 + \gamma^2 \sin^2 \theta} \cos^2 \theta d \psi^2 + \f{\rho^2 }{\rho^2 + \gamma^2 \sin^2 \theta} \sin^2 \theta d \phi^2 \Bigg] \label{Eqn:InnerMetricHalf&Half}
\eea

\subsection{Spectral flow}

We do a spectral flow by $(2n)$ units in both the sectors. This is accomplished by
\be
\psi \rightarrow \psi + 2 n \tau,  \qquad \phi \rightarrow \phi + 2 n \varphi
\ee
and the metric is then
\bea
ds^2 = \sqrt{Q_1 Q_5} \Bigg[ - (\rho^2 + \gamma^2 \sin^2 \theta) ( d\tau^2 - d \varphi^2)+ \f{d \rho^2}{\rho^2 + \gamma^2}~~~~~~~~~~~~~~~~~~~~~~~~~~~~~~~~~~&&{} \nn
+ d \theta^2 + \f{\rho^2 + \gamma^2}{\rho^2 + \gamma^2 \sin^2 \theta} \cos^2 \theta (d \psi+(2n) d\tau)^2 + \f{\rho^2 }{\rho^2 + \gamma^2 \sin^2 \theta} \sin^2 \theta (d \phi+ (2n) d\varphi)^2& \Bigg] & \label{Eqn:InnerOrbifoldMetricHalfNHalfSF}
\eea

\subsection{Ergoregion}

Let us now find the ergoregion of this geometry.
The ergoregion is given by
\be
g_{\tau \tau} >0
\ee
For the case of the metric \bref{Eqn:InnerOrbifoldMetricHalfNHalfSF} this is given by
\be
 -( \rho^2 + \gamma^2 \sin^2 \theta)^2 + (2n)^2 (\rho^2+\gamma^2) \cos^2 \theta >0
\ee
which gives
\be
0< \rho < \sqrt{ (\cos^2 \theta(n + \sqrt{n^2+ \gamma^2}))^2 - \gamma^2}
\label{ergo2}
\ee

\subsection{Ergoregion for large spectral flow}

We have done a spectral flow by $2n$ units for each of the left and right sectors. Suppose we take $n\gg 1$.  In this limit we find that the ergoregion (\ref{ergo2}) becomes
\be
0 < \rho < 2n \cos \theta
\label{ergofinal}
\ee
Let us compare this to the ergoregion for the geometries of \cite{ross}, given in eq. \bref{Eqn:RossErgoregion}. Note that
\be
m=2nk+1
\ee
For large $n$ we find that the ergoregion (\ref{Eqn:RossErgoregion}) becomes
\be
0 < \rho < 2n \cos \theta
\ee
Thus we see that in the limit of large spectral flow the ergoregion (\ref{ergofinal}) agrees with the ergoregion of the geometries of \cite{ross}. We can trace this agreement to the fact that spectral flow adds angular momentum, and in the limit of large spectral flow the dominant part of the angular momentum comes from this spectral flow. Ergoregions arise from angular momentum, and thus tend to agree when the spectral flow is made large, regardless of the extremal geometry that we started with.

\section{Discussion}

To resolve the information paradox it is crucial to understand how radiation emerges from a black hole. The computations of \cite{ross,myers,cm1,cm2} have shown that for a special family of microstates, we have an ergoregion rather than a horizon, and the radiation arises as pair creation in this ergoregion. One  member of the pair stays in the ergoregion, while the other escapes to infinity.\footnote{For other computations with ergoregions, see \cite{ashtekar,friedman,cominsschutz,kang}. A nice application to superradiance in the string theory D1-D5 context can be found in \cite{emparan}.}

\begin{figure}[htbp] 
   \begin{center}
   \includegraphics[width=2in]{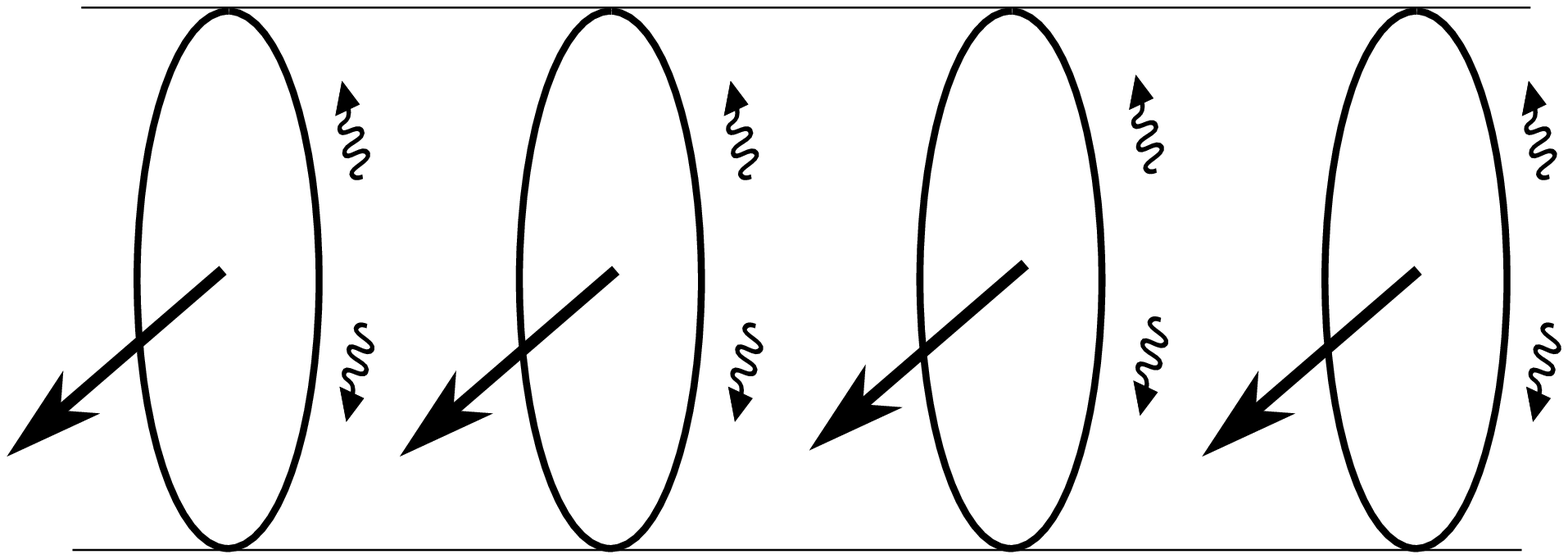}   \hspace{1.5truecm}
   \includegraphics[width=2in]{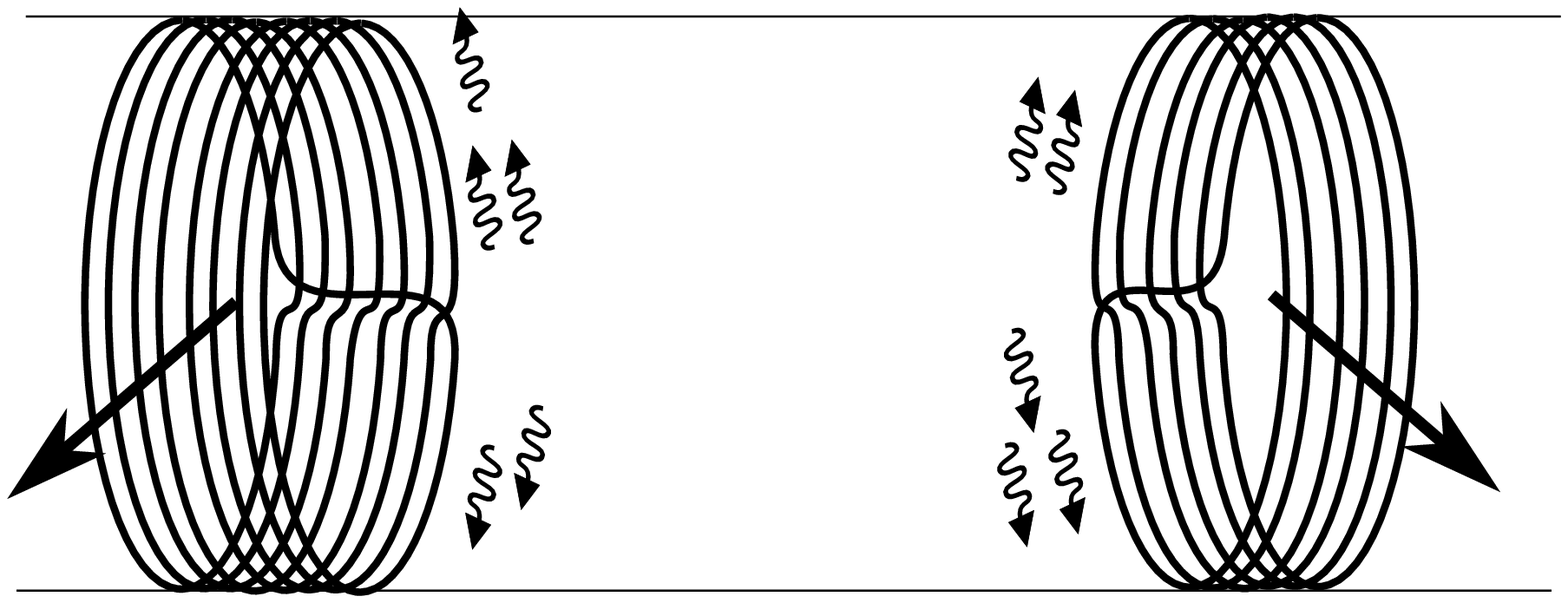} \\
   \vspace{.2truecm}
   \end{center}
    \hspace{4.2truecm} (a)\hspace{6.4truecm}(b)
      \vspace{1truecm}
   \begin{center}
   \includegraphics[width=2in]{SpectralFlowExcitedMany.eps} \hspace{1.5truecm}
   \includegraphics[width=2in]{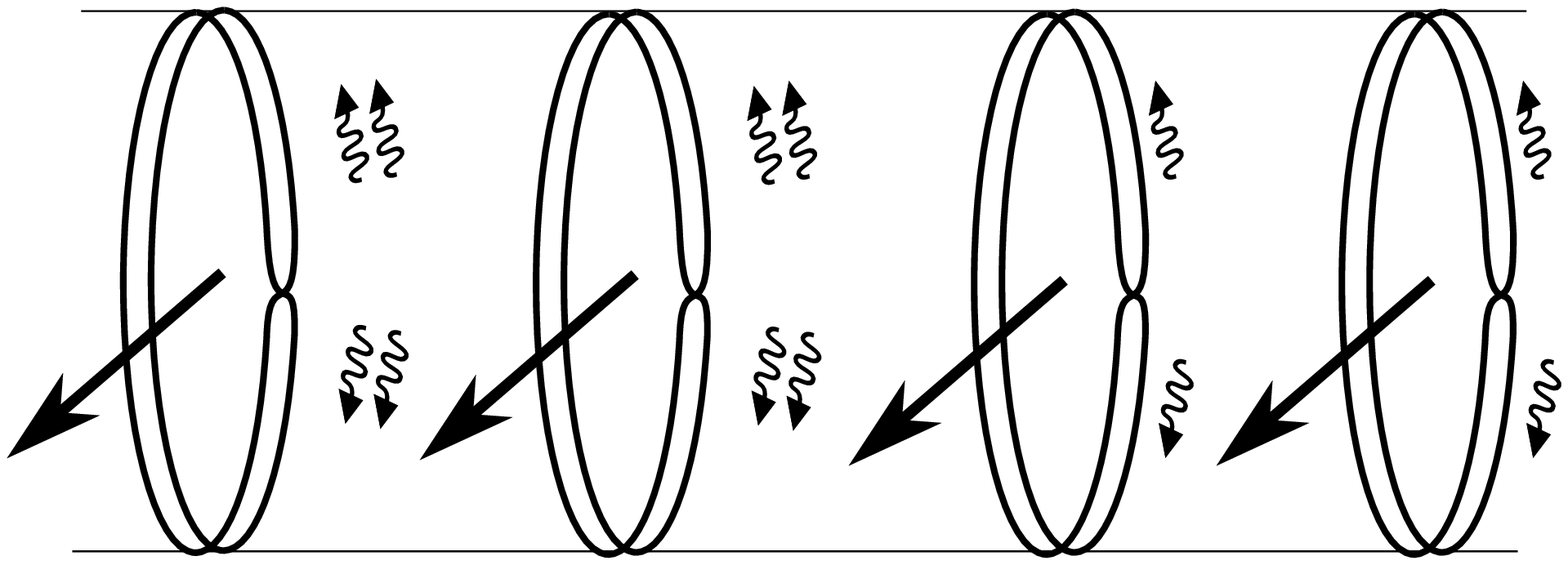} \\
   \vspace{.2truecm}
   \end{center}
     \hspace{4.2truecm} (c)\hspace{6.4truecm}(d)
   \caption{(a) The CFT state with winding number $k=1$ \cite{cm1} (b) The generic CFT state dual to the black hole (c) CFT states with $k\ne 1$ (d) The CFT state for the geometry (\ref{Eqn:InnerOrbifoldMetricHalfNHalfSF}); the component strings are not all the same. }
   \label{fig:discussion}
\end{figure}

The CFT state considered in \cite{cm1} is depicted in figure \ref{fig:discussion} (a). All component strings have winding number $k=1$ and all spins are aligned. The {\it generic} state of the non-extremal D1-D5-P hole is depicted in figure \ref{fig:discussion} (b). There are multiwound component strings, with different lengths, base spins and excitations.

In the present paper we have considered CFT states where the component strings have winding $k\ne 1$ (figure \ref{fig:discussion} (c)). Again we obtained precise agreement between the CFT emission rate and the rate of pair creation in the ergoregion.

We then considered emission from more general microstates with ergoregions where the wave-equation could not be solved explicitly in closed form. We developed a way of writing down the emission in terms of a function $F(\omega)\equiv \left({F_1\over F_2}\right)(\omega)$, which described the large distance fall-off behavior of the scalar field wavefunction in the `cap' of the geometry. Finally, we made another family of microstates for the non-extremal D1-D5 system by spectral flowing a set of extremal geometries. While we have not given the construction of the dual CFT states in this paper, one can carry out this construction following the methods in \cite{lm4}. The CFT state has component strings that are {\it  not} all the same. The winding number distribution was peaked around the value $k$ in eq. (\ref{gamma2k}), but  the excitations on half the component strings are of one kind and on the other half are of a different kind (figure \ref{fig:discussion} (d)).

These studies take us further along our goal of arguing that black holes do not have a traditional horizon, which has no `information' in its vicinity. Instead, microstates of black holes have their information distributed all over a horizon sized ball, and information emerges from this ball just as it would from a piece of coal. The constructions of \cite{cm1} and the present paper give explicit examples of this process for simple microstates. These microstates are simple in that we have taken many  component strings to be in the same state; this makes the dual geometry very `classical'. As we reduce the number of component strings of each type (i.e. spread the total winding over many different types of component strings) the state develops more quantum fluctuations. The generic state has very few component strings of any given type, and is thus a  `quantum fuzzball', but we expect that the essential physics will be obtained as a logical limit where we move from the simple classical states to progressively more quantum ones.

\section*{Acknowledgments}
\setcounter{equation}{0}

We thank Steve Avery and Stefano Giusto for many helpful comments. This work was supported in part by DOE grant DE-FG02-91ER-40690.

\end{document}